\documentclass[sigconf]{acmart}

\AtBeginDocument{%
  }

\setcopyright{acmlicensed}
\copyrightyear{2025}
\acmYear{2025}
\acmDOI{XXXXXXX.XXXXXXX}
\acmConference[CIKM '25]{The 34th ACM International
Conference on Information and Knowledge Management}{November 10--14,
  2025}{Ceox, Seoul, Korea}
\acmISBN{978-1-4503-XXXX-X/2018/06}

\newif\ifstatus
\statustrue
\statusfalse

\begin{document}

\title{State \& Geopolitical Censorship on Twitter (X): Detection \& Impact Analysis of Withheld Content}

\author{Yusuf M\"{u}cahit \c{C}etinkaya}
\email{v1ymucah@ed.ac.uk}
\orcid{0000-0001-5338-750X}
\affiliation{%
  \institution{University of Edinburgh}
  \city{Edinburgh}
  \country{United Kingdom}
}

\author{Tuğrulcan Elmas}
\email{telmas@ed.ac.uk}
\orcid{0000-0002-4305-1479}
\affiliation{%
  \institution{University of Edinburgh}
  \city{Edinburgh}
  \country{United Kingdom}
}

\renewcommand{\shortauthors}{\c{C}etinkaya and Elmas}

\begin{abstract}
   State and geopolitical censorship on Twitter, now X, has been turning into a routine, raising concerns about the boundaries between criminal content and freedom of speech. One such censorship practice, withholding content in a particular state has renewed attention due to Elon Musk's apparent willingness to comply with state demands. In this study, we present the first quantitative analysis of the impact of state censorship by withholding on social media using a dataset in which two prominent patterns emerged: Russian accounts censored in the EU for spreading state-sponsored narratives, and Turkish accounts blocked within Turkey for promoting militant propaganda. We find that censorship has little impact on posting frequency but significantly reduces likes and retweets by 25\%, and follower growth by 90\%—especially when the censored region aligns with the account's primary audience. Meanwhile, some Russian accounts continue to experience growth as their audience is outside the withholding jurisdictions. We develop a user-level binary classifier with a transformer backbone and temporal aggregation strategies, aiming to predict whether an account is likely to be withheld. Through an ablation study, we find that tweet content is the primary signal in predicting censorship, while tweet metadata and profile features contribute marginally. Our best model achieves an F1 score of 0.73 and an AUC of 0.83. This work informs debates on platform governance, free speech, and digital repression.

\end{abstract}





\maketitle

\section{Introduction}
State and geopolitical censorship on social media platforms like Twitter (X) has become a critical and growing concern. As governments increasingly leverage digital spaces to control narratives and suppress dissent, the line between legitimate content moderation and political suppression continues to blur. This complexity is further deepened by shifting platform policies—often influenced by political and geopolitical pressures—that determine how and when content is withheld. For instance, state-affiliated account networks spreading propaganda (troll) were suspended under the Civic Integrity project by Twitter, and the account data were shared until 2022. Twitter previously labeled state-affiliated accounts, such as the notable Russian news agency Russia Today, and soft-banned them by excluding them from recommendation algorithms and decreasing their reach. These practices are now discontinued under Elon Musk's ownership~\cite{james2023putins}. 

Our focus in this paper is withholding content, a common censorship practice in which the content does not violate the platform's terms of service, but the platform censors it by withholding it upon a legal request in the requesting jurisdiction, instead of suspending. How much Twitter accommodates or does not accommodate these requests is still controversial~\cite{lamensch2024digital}. Elon Musk's ownership has renewed public attention to the withholding content, as it contradicts his proclaimed vision of absolute free speech~\cite{rohlinger2023does,hickey2023auditing,robinson2024moderated}. This contradiction became most evident in March 2025 when X withheld İstanbul mayor Ekrem İmamoğlu’s account of 10 million followers and those of his supporters including university students, after his presidential candidacy was announced by the main opposition party in Turkey, intensifying debates over freedom of speech on X~\cite{orla2025turkey}.

Research on social media censorship has explored government takedown requests~\cite{tanash2015known, varol2016spatiotemporal, voinea2022taking, caesmann2024censorship, gosztonyi2025public}, and several studies have examined the role and effects of automated moderation in account suspension or shadow bans~\cite{volkova2017identifying, chowdhury2021examining, toraman2022blacklivesmatter, pierri2023does, elmas2023analyzing, jaidka2023silenced, diskin2024use, ccetinkaya2024towards}. Yet, much of this work has relied on event-based or snapshot analyses, often isolating specific campaigns or focusing solely on individual tweets rather than users' longitudinal behavior. To the best of our knowledge, our study is the first to quantify the sustained impact of state-imposed censorship on user activity and engagement over time.

We use two datasets in our analysis. The first is a preexisting dataset containing censored content up to 2021, including takedowns in Germany, France, Russia, India, and Turkey. Among these, Turkey stands out due to the prevalence of accounts spreading militant propaganda~\cite{elmas2021dataset}. The second is a new dataset we collected by focusing on the EU's decision to censor Russian propaganda accounts over the Russia-Ukraine war~\cite{james2022twitter, geissler2023russian}. These cases offer clear and high-impact examples of how geopolitical tensions shape social media governance. Our research questions are the following:

\vspace{.1cm}
\noindent\textbf{RQ1:} What is the impact of censorship on users' activity and the engagements they receive?

\vspace{.1cm}
\noindent\textbf{RQ2:} Is censorship predictable based on content and profile signals?

Our main contribution is the first systematic analysis of state-specific withholding on social media—capturing not just whether content is censored, but how users alter their posting behavior and engagement patterns when their tweets are selectively suppressed. We complement the analysis by a predictive task that models censorship at the user level by aggregating sequences of tweets, instead of treating each tweet in isolation. This approach provides a holistic view of how withholding unfolds over time, and it demonstrates, via an ablation study, that raw tweet content alone is the dominant predictive signal for withholding decisions, showing that metadata and profile features contribute little.

\section{Data}
\noindent\textbf{Censorships until 2021:} The first dataset is collected using the Internet Archive's Twitter Stream Grab, which contains 1\% of all tweets between September 2011 and June 2020, from which all tweets with a non-empty "withheld in countries" field are extracted, yielding 583,437 censored tweets from 155,715 unique users~\cite{elmas2021dataset}. Fully censored accounts are identified via the Twitter User Lookup API and an inference heuristic, resulting in 4,301 entirely withheld users. To support analysis, about 22 million non-censored tweets (including retweets) from users with at least one censored tweet are also collected. 98\% of the censorships occur in Turkey, Germany, France, India, and Russia. 

\noindent\textbf{2022 EU Censorship of Russian Propaganda Accounts:} The second dataset is collected using the 10\% Decahose Twitter API between 2022 and 2023. We identified the accounts that were blocked in all EU countries and the UK through the ``withheld in countries" field provided by the Twitter API, which are all Russian propaganda accounts as this was the only such instance where these countries unanimously censored accounts. We identified 38 accounts including Russia Today, Sputnik, and Ruptly. We collected all tweets by them, resulting in 148,037 tweets before and 123,685 tweets after withholding in March 2022.

\section{RQ1: Impact Analysis}
\noindent\textbf{Methodology:} We analyze the impact of censorship on the accounts' behavior and engagements by comparing the respective measures before and after the withholding event. We repeat this comparison for a control group. We employ the Wilcoxon signed‐rank test to assess whether the changes before and after, for each measure are statistically significant. The Wilcoxon signed‐rank test compares paired pre‐ and post‐treatment (e.g., withholding event) measurements by ranking the absolute differences and assessing whether their median shift differs from zero~\cite{wilcoxon1992individual}.

Since we do not have the exact time each account was completely withheld, we use the date of the first withheld tweet of each account that has all tweets after that time are withheld as a proxy for the time the account was withheld.

We measure the impact within a time window of \(W\) (e.g., 30 days). For each window \(W\), we use the withheld time \( t_i \) to compute the pre-treatment interval \(\bigl[ t_i - W,\, t_i - 1 \bigr]\) and the post-treatment interval \(\bigl[ t_i,\, t_i + W \bigr]\). We merge each account's corresponding tweet data in each interval. We then aggregate per‐account metrics over each interval.

We compute these metrics as follows: for an interval \(win\) (pre- or post-) and an account \(i\), we compute the median of the daily retweets the account receives in the interval \((\mathrm{med\_retweets}_{i,\mathrm{win}})\), the median of the daily likes the account receives in the interval \((\mathrm{med\_likes}_{i,\mathrm{win}})\), the average number of tweets the account posted per day in the interval \((\mathrm{posts\_per\_day}_{i,\mathrm{win}})\), and the average daily follower gain in the interval \((\mathrm{avg\_follower\_gain}_{i,\mathrm{win}})\).

Let \(D_{i,M}^{(W)}\) denote the paired difference for a metric \(M\), window \(W\), and a user \(i\), defined as the value of \(M\) in the post‐withholding window minus the value of \(M\) in the pre‐withholding window. We then apply the Wilcoxon signed‐rank test to the collection of paired differences \(D_{i,M}^{(W)}\).

\noindent\textbf{Experiments:}
We evaluate multiple symmetric windows around each withholding event by iterating over \(W \in \{30,45,60,75,90,120,\\150\}\) days. Any account with fewer than two tweets in either interval is excluded. The surviving authors in the first window are used for subsequent windows (\(W=45,60,\dots\)) to use an identical cohort of withheld accounts. To construct a matched control group, for each investigated withheld author \(a\), we compute their mean follower count. We then sample one account \(c\) uniformly at random from the accounts that do not have any withheld tweets, whose mean follower count lies within \(\pm 10\%\) of \(a\)'s follower count. The resulting list has 197 accounts that meet the conditions mentioned above and the same size control group as the withheld cohort, using the midpoint of their activities as synthetic withheld times using the first dataset that is until 2021. Additionally, we apply the same analysis to the Russian state-affiliated accounts without building a control group.

\begin{table}[t]
  \centering
  \caption{Pre- vs. post-withholding engagement and tweets posted per day for the withheld and control groups across different windows (W). \\
  \textnormal{* $p<0.05$ for Pre vs. Post within the same group (Wilcoxon test).}}
  \label{tab:engagement}
  \begin{tabular}{l
                  |r@{\,}r@{\;\;}|r@{\,}r@{\;\;}   
                  |r@{\,}r@{\;\;}|r@{\,}r}         
    \toprule
    
      & \multicolumn{4}{c|}{\textbf{Withheld}} 
      & \multicolumn{4}{c}{\textbf{Control}} \\
    \cmidrule(lr){2-5} \cmidrule(lr){6-9}
    \textbf{W}  & Pre & Post 
      & Pre & Post 
      & Pre & Post 
      & Pre & Post \\
    \midrule
      & \multicolumn{2}{c|}{RTs} 
      & \multicolumn{2}{c|}{Likes} 
      & \multicolumn{2}{c|}{RTs} 
      & \multicolumn{2}{c}{Likes} \\
    30 
      & 24 & 18* 
      & 20 & 15* 
      & 40 & 40 
      & 44 & 50 \\
    45 
      & 22 & 17* 
      & 19 & 14* 
      & 40 & 42 
      & 42 & 53 \\
    60 
      & 22 & 16* 
      & 19 & 13* 
      & 36 & 39 
      & 39 & 50 \\
    75 
      & 21 & 16* 
      & 17 & 13* 
      & 35 & 41 
      & 38 & 52 \\
    90 
      & 21 & 16* 
      & 17 & 13* 
      & 34 & 39 
      & 37 & 50 \\
    120
      & 20 & 15* 
      & 16 & 13* 
      & 32 & 41 
      & 35 & 50 \\
    150
      & 20 & 15* 
      & 16 & 13* 
      & 31 & 39 
      & 35 & 48 \\
    \midrule
    & Pre & Post 
      & Pre & Post 
      & Pre & Post 
      & Pre & Post \\
    \midrule
      & \multicolumn{2}{c}{Follwr. $\Delta$} 
      & \multicolumn{2}{c|}{Status $\Delta$} 
      & \multicolumn{2}{c}{Follwr. $\Delta$} 
      & \multicolumn{2}{c}{Status $\Delta$} \\
    30 
      & 140 & 15* 
      & 86 & 54* 
      & 65 & 56* 
      & 62 & 56* \\
    45 
      & 137 & 18* 
      & 84 & 59* 
      & 66 & 50* 
      & 62 & 59* \\
    60 
      & 136 & 16* 
      & 80 & 59* 
      & 62 & 46* 
      & 59 & 55* \\
    75 
      & 135 & 14* 
      & 75 & 53* 
      & 63 & 44* 
      & 59 & 55* \\
    90 
      & 136 & 13* 
      & 74 & 54* 
      & 60 & 43* 
      & 57 & 53* \\
    120
      & 136 & 12* 
      & 32 & 54* 
      & 58 & 40* 
      & 55 & 51* \\
    150
      & 136 & 10* 
      & 40 & 55* 
      & 58 & 38* 
      & 55 & 49* \\
    \bottomrule
  \end{tabular}
\end{table}

\noindent\textbf{Results:} The analysis reveals a consistent and statistically significant decline in user engagement metrics following the withholding events, as summarized in Table~\ref{tab:engagement}. The control group's engagement averages do not change, and they are not statistically significant, where this pattern is observed across all examined window sizes. Withholding actions effectively reduce user interactions and visibility on the platform. The results show about a 90\% decrease in follower gain per day after withholding. The control group's follower gain change is also statistically significant, but it is not more than 15\%. This can be explained by users reaching saturation~\cite{jain2025follower}. The decrease in the status count posted per day, on the other hand, is not as sharp, even though it is statistically significant.

\begin{figure}[h]
  \centering
  \includegraphics[width=\linewidth]{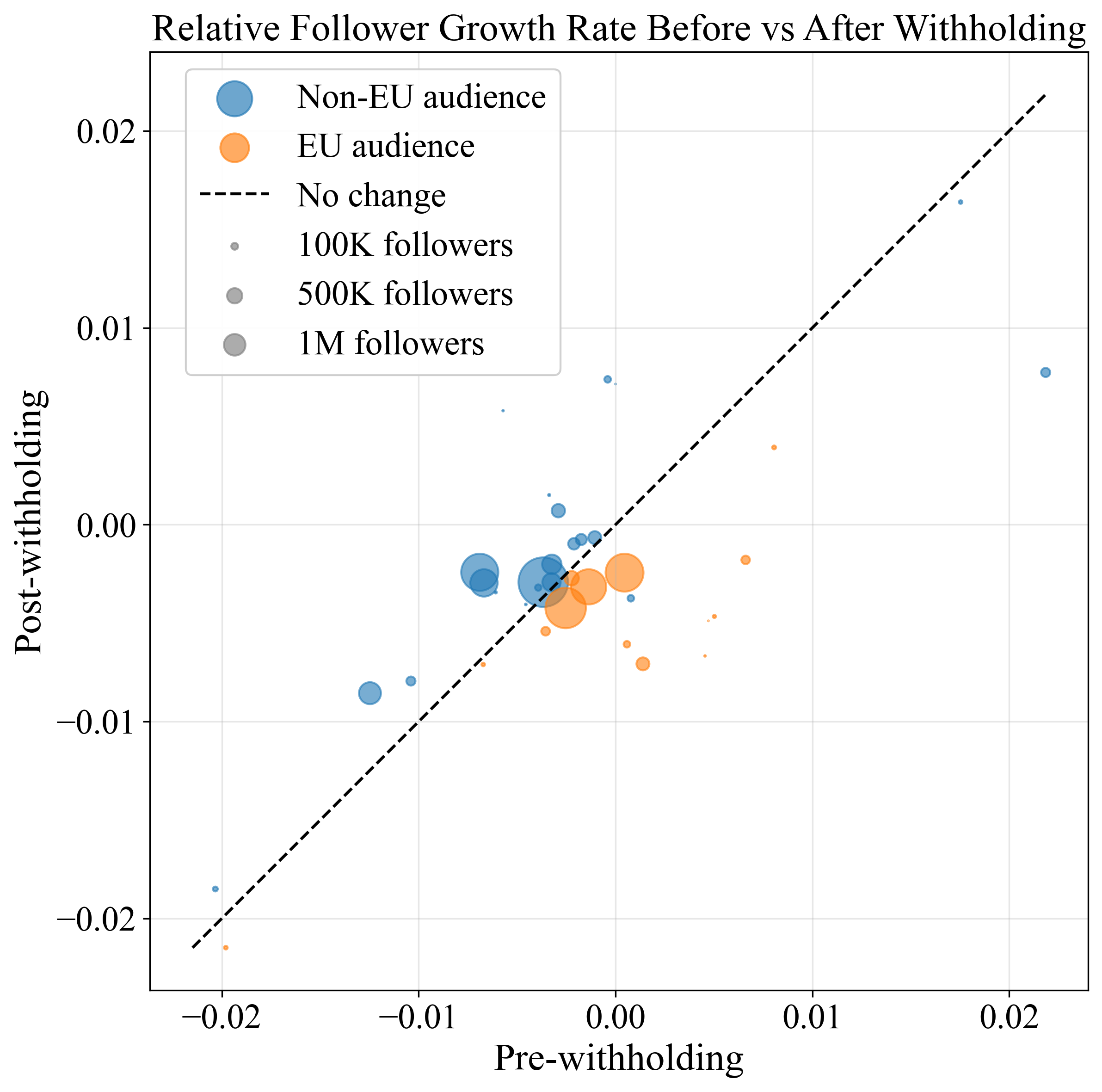}
  \vspace{-.5cm}
  \caption{Withheld Russian accounts' relative daily follower gains ratio to follower sizes with their target audiences.}
  \label{fig:russian-accounts}
\end{figure}

The analysis of Russian accounts withheld in the EU did not yield statistically significant results, necessitating a more in-depth analysis. The scatter plot in Figure~\ref{fig:russian-accounts} illustrates the relative follower growth rates before and after the EU sanctions for 38 Russian accounts. Bubble sizes denote the initial follower count of each account. While most EU-audience accounts lie below the no-change line, indicating a decrease in follower growth after withholding, some exceptions emerge. For example, rentvchannel and Sputnik Serbia, despite targeting non-EU audiences, exhibit a positive growth trend before and after withholding, though the growth rate diminishes post-withholding. Similarly, RTonline\_ar experienced positive follower growth before withholding but a negative rate afterward. Notably, Sputnik Brazil (0.06→0.03) and Ruptly (-0.01→0.09) are not shown within the plotting window.

The trends differ based on the preexisting trajectory of the accounts' follower growth: accounts already experiencing follower loss see a reduction in the rate of loss, while accounts with a prior gain trend show a diminished rate of gain. This pattern may be attributable to the removal of the unfollow button following the withholding action. Further grouping of engagement and follower behavior changes among these Russian accounts reveals variations in their target audiences. Notably, the analysis of withheld Russian accounts with EU-based audiences aligns with the findings for the accounts in the first dataset.

\section{RQ2: Classification}

\begin{figure*}[t]
  \centering
  \includegraphics[width=\linewidth]{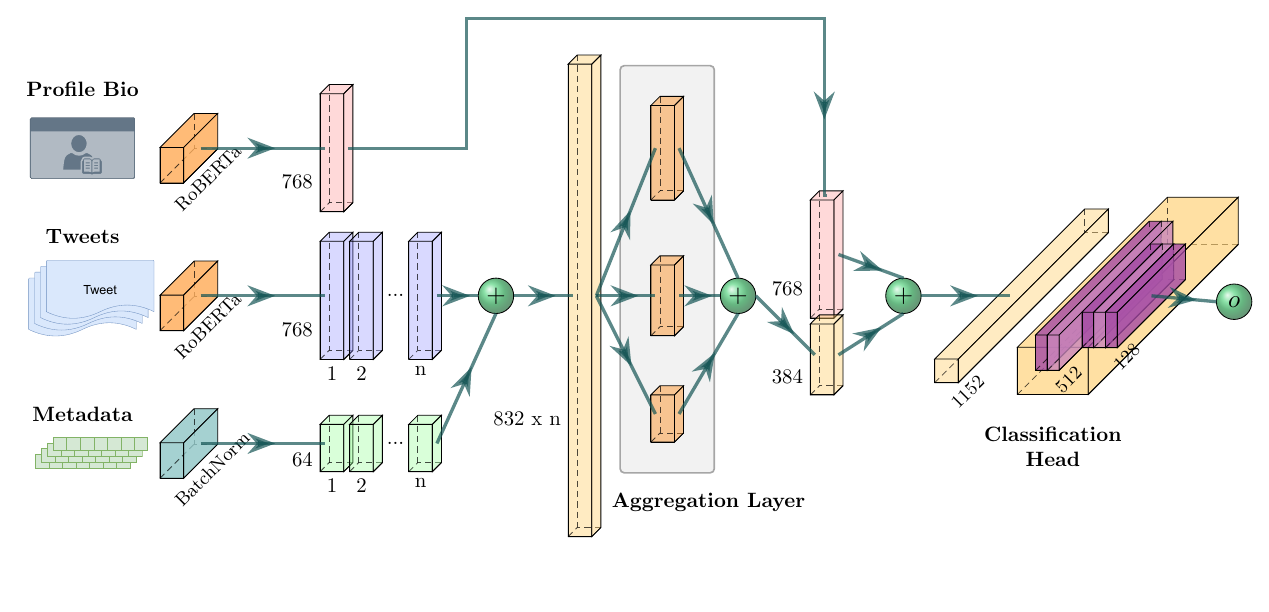}
  \vspace{-1.2cm}
  
  \caption{The architecture of our TwCensorNet for predicting censorship on an account.}
  \label{fig:architecture}
\end{figure*}

\noindent\textbf{Methodology:} We introduce a model named TwCensorNet for a \emph{user-level} binary classification task: given up to 50 tweets and associated metadata for a Twitter account, we predict whether the account could be withheld. Figure~\ref{fig:architecture} shows the model architecture. Each user $u$ is represented by a set of tweets $\{\mathbf{x}_{u,i}\}$ and optional tweet-level metadata $\{\mathbf{m}_{u,i}\}$ (retweet count, like count, follower/friend counts at posting), alongside profile features $\mathbf{p}_u$ (concatenated screen name (e.g., @realDonaldTrump), display name (e.g., Donald Trump), profile description). The model outputs $\hat{y}_u = \Pr(y_u = 1 \mid \{\mathbf{x}_{u,i},\mathbf{m}_{u,i}\}, \mathbf{p}_u, \mathbf{q}_u)$ via a transformer backbone and downstream aggregation.

Each tweet $t_{u,i}$ is encoded by the \texttt{twitter-roberta-base} \cite{barbieri2020tweeteval} model, trained on \(\sim\)~58M tweets on top of the original RoBERTa-base checkpoint, yielding a $768$-dimensional embedding $\mathbf{h}_{u,i}$. If tweet-level metadata is enabled, $\mathbf{m}_{u,i}\in\mathbb{R}^5$ is batch-normalized and projected to a $64$-dimensional hidden space via BatchNorm → Linear → ReLU → BatchNorm, producing $\hat{\mathbf{m}}_{u,i}\in\mathbb{R}^{64}$. The per-tweet representation becomes $\mathbf{r}_{u,i} = [\mathbf{h}_{u,i}\Vert\hat{\mathbf{m}}_{u,i}] \in \mathbb{R}^{832}$; otherwise, $\mathbf{r}_{u,i} = \mathbf{h}_{u,i}$ ($768$-dimensional).

To collapse the variable-length sequence $\{\mathbf{r}_{u,1},\dots,\mathbf{r}_{u,N_u}\}$ into a fixed-size vector $\mathbf{a}_u$, TwCensorNet supports five aggregators: (1) \emph{Mean Pooling} ($\tfrac{1}{N_u}\sum_i\mathbf{r}_{u,i}$), (2) \emph{Max Pooling} ($\max_i \mathbf{r}_{u,i}$), (3) \emph{Attention} ($e_{u,i}=w_a^\top\mathbf{r}_{u,i}$, $\alpha_{u,i}=\mathrm{softmax}(e_{u,i})$, $\sum_i\alpha_{u,i}\mathbf{r}_{u,i}$), (4) \emph{Convolutional} (1D convolutions with kernel sizes $\{3,5,7\}$, ReLU, temporal max-pool, concatenated to $3\times128=384$ dims), and (5) \emph{BiLSTM} (bidirectional LSTM with hidden size $384$ per direction, final hidden state of size $768$).

Optionally, concatenated profile text $\mathbf{p}_u$, encoded via the same RoBERTa backbone to a $768$-dimensional vector, is fused to $\mathbf{a}_u$, yielding $\mathbf{u}_u\in\mathbb{R}^{d_{\text{agg}}+768}$. The classifier head is a three-layer MLP: Linear($d_{\text{in}}\to512$) → ReLU → Dropout(0.3) → Linear($512\to128$) → ReLU → Dropout(0.3) → Linear($128\to1$) → Sigmoid.

\noindent\textbf{Experiments:} 
To construct the dataset of users with similar interests, we add the users who retweeted at least one withheld account and also posted at least one tweet of their own to the list of withheld accounts. The dataset contains 318,040 tweets belonging to 2{,}227 withheld and 3,156 non-withheld accounts. The Russian-affiliated media dataset is not included since the classifier would then be predicting whether an account is a Russian news account or not. These users are then split into training, validation, and test sets in a 70\%/15\%/15\% ratio, ensuring no overlap of users across splits. Within each split, up to 50 tweets per user are retrieved, randomly sampled if more are available, and labels reflected whether the user is withheld or not. The resources are publicly available on Github at \href{https://github.com/tweetpie/twitter-censorship}{tweetpie/twitter-censorship} repository.

For each model, performance on the validation set is measured by ROC–AUC. Separately, we compute F1 by thresholding the sigmoid output: for each architecture, the threshold $\tau$ is chosen to maximize F1 on validation. Specifically, after obtaining $\hat{y}_u\in(0,1)$ for each user $u$, binary predictions $\hat{y}_u^{(\tau)} = \mathbb{I}[\hat{y}_u > \tau]$ are compared to ground truth to compute precision, recall, and F1 via standard definitions. The optimal $\tau$ per model is the value yielding the highest F1 on the validation split. The model checkpoint with the highest validation ROC–AUC is selected for final testing.

All experiments use identical hyperparameters unless noted: convolutional aggregators employ kernel sizes $k\in\{3,5,7\}$ with 128 filters each, a BiLSTM hidden size of 384 per direction, and the classifier MLP has layer sizes [512,128] with dropout $p=0.3$. Optimization is via AdamW at a learning rate of $2\times10^{-3}$, batch size 16, for up to 20 epochs. Early stopping on validation ROC–AUC uses a minimum delta of 0.001 and a patience of 5 epochs.

\noindent\textbf{Results:} Table~\ref{tab:model_performances} summarizes the performance of various models on the task of predicting withholding status. Both the convolutional and Bi-LSTM architectures achieve comparable results in terms of ROC-AUC and test F1 scores, with the convolutional model demonstrating a slight performance advantage overall. The optimal thresholds for maximizing the F1 scores of most models fall between 0.3 and 0.4, suggesting a mild tendency toward predicting accounts as not withheld.

\begin{table}[ht]
  \caption{Effectiveness of aggregation methods, ablation analyses, and decision thresholds in TwCensorNet.}
  \label{tab:model_performances}
  \centering
  \begin{tabular}{c|c|c|l}
    \toprule
    \textbf{ROC–AUC} & \textbf{Test F1} & \textbf{Threshold} & \textbf{Variant} \\ 
    \midrule
    0.75            & 0.65           & 0.50              & Mean \\ 
    0.77            & 0.67           & 0.30              & Max \\ 
    0.78            & 0.69           & 0.30              & Attention \\ 
    0.82            & 0.71           & 0.30              & BiLSTM \\ 
    0.82            & 0.71           & 0.40              & Conv \\ 
    0.82            & \textbf{0.73}  & 0.40              & Conv + Meta \\ 
    0.82            & \textbf{0.73}  & 0.30              & Conv + Profile \\ 
    \textbf{0.83}   & \textbf{0.73}  & 0.30              & Conv + Meta + Profile \\ 
    \bottomrule
  \end{tabular}
\end{table}

Furthermore, incorporating either metadata or profile features generally enhances the model's predictive performance relative to using tweet content alone. However, the simultaneous addition of both metadata and profile features does not yield substantial further improvements. While these additional features contribute to performance gains individually, tweet content remains the most informative signal for predicting withholding status.

\section{Conclusion and Future Work}

Our study reveals several key insights into the dynamics of state and geopolitical censorship on Twitter (X). The analysis of \textit{RQ1} demonstrates that, although censorship events typically lead to a modest decline in posting frequency, they do not appear to fundamentally suppress the underlying motivation of the censored accounts. The accounts continue to engage with their audiences, albeit at a slightly reduced rate.

Meanwhile, the results underscore the severe impact of censorship on an account's visibility and reach, particularly when the censorship-enforcing government coincides with the account's primary audience. Engagement metrics and follower growth rates show pronounced declines in such cases, indicating that censorship can effectively curtail the influence of targeted accounts. Conversely, accounts with audiences outside the enforcing government's jurisdiction experience more muted effects.

In addressing \textit{RQ2}, our findings highlight that censorship decisions are to a significant extent predictable. The results emphasize that social media censorship has a profound impact on the visibility of these accounts, underscoring the critical need for governments to carefully navigate the delicate boundary between combating criminal activity and safeguarding freedom of expression.

Future work may explore several promising avenues. It is noteworthy that an account's size, as measured by follower count, does not directly determine its susceptibility to censorship. Examining shifts in follower numbers, particularly in the aftermath of censorship events, offers a valuable line of inquiry.

Conducting a content analysis could reveal themes and rhetorical patterns in censored posts, clarifying what types of discourse are most likely to draw government intervention. Investigating the network graph structures of involved accounts might uncover whether coordinated behaviors, such as clustered retweeting or shared content, precede or coincide with censorship actions.

\noindent\textbf{Limitations:} As we rely on 1\% and 10\% samples rather than the full Twitter ``firehose,'' our analysis is limited to subsets of the data. This affects our construction of matched control accounts, which are based solely on follower counts and may differ from censored accounts in content, language, or activity patterns. Additionally, our engagement metrics do not capture other important signals of influence, such as view counts or quote tweets. 

\noindent\textbf{Ethics Statement:} While our model could be misused by others for proactive unethical censorship, its purpose is to enhance transparency, accountability, and public understanding of platform governance. It should not be applied to justify suppression of lawful expression, except in cases of criminal content where removal is both legally and ethically warranted.

\section*{GenAI Usage Disclosure}
The authors wish to disclose that GenAI tools were employed solely for the purpose of proofreading and minor editorial enhancement of this manuscript.

\bibliographystyle{ACM-Reference-Format}
\bibliography{references}


\begin{thebibliography}{25}


\ifx \showCODEN    \undefined \def \showCODEN     #1{\unskip}     \fi
\ifx \showISBNx    \undefined \def \showISBNx     #1{\unskip}     \fi
\ifx \showISBNxiii \undefined \def \showISBNxiii  #1{\unskip}     \fi
\ifx \showISSN     \undefined \def \showISSN      #1{\unskip}     \fi
\ifx \showLCCN     \undefined \def \showLCCN      #1{\unskip}     \fi
\ifx \shownote     \undefined \def \shownote      #1{#1}          \fi
\ifx \showarticletitle \undefined \def \showarticletitle #1{#1}   \fi
\ifx \showURL      \undefined \def \showURL       {\relax}        \fi
\providecommand\bibfield[2]{#2}
\providecommand\bibinfo[2]{#2}
\providecommand\natexlab[1]{#1}
\providecommand\showeprint[2][]{arXiv:#2}

\bibitem[Barbieri et~al\mbox{.}(2020)]%
        {barbieri2020tweeteval}
\bibfield{author}{\bibinfo{person}{Francesco Barbieri}, \bibinfo{person}{Jose
  Camacho-Collados}, \bibinfo{person}{Leonardo Neves}, {and}
  \bibinfo{person}{Luis Espinosa-Anke}.} \bibinfo{year}{2020}\natexlab{}.
\newblock \showarticletitle{TweetEval: Unified benchmark and comparative
  evaluation for tweet classification}.
\newblock \bibinfo{journal}{\emph{arXiv preprint arXiv:2010.12421}}
  (\bibinfo{year}{2020}).
\newblock


\bibitem[Caesmann et~al\mbox{.}(2024)]%
        {caesmann2024censorship}
\bibfield{author}{\bibinfo{person}{Marcel Caesmann}, \bibinfo{person}{Janis
  Goldzycher}, \bibinfo{person}{Matteo Grigoletto}, {and}
  \bibinfo{person}{Lorenz Gschwent}.} \bibinfo{year}{2024}\natexlab{}.
\newblock \showarticletitle{Censorship in democracy}.
\newblock \bibinfo{journal}{\emph{arXiv preprint arXiv:2406.03393}}
  (\bibinfo{year}{2024}).
\newblock


\bibitem[{\c{C}}etinkaya et~al\mbox{.}(2024)]%
        {ccetinkaya2024towards}
\bibfield{author}{\bibinfo{person}{Yusuf~M{\"u}cahit {\c{C}}etinkaya},
  \bibinfo{person}{Yeonjung Lee}, \bibinfo{person}{Emre K{\"u}lah},
  \bibinfo{person}{{\.I}smail~Hakk{\i} Toroslu}, \bibinfo{person}{Michael~A
  Cowan}, {and} \bibinfo{person}{Hasan Davulcu}.}
  \bibinfo{year}{2024}\natexlab{}.
\newblock \showarticletitle{Towards a Programmable Humanizing AI through
  Scalable Stance-Directed Architecture}.
\newblock \bibinfo{journal}{\emph{IEEE Internet Computing}}
  (\bibinfo{year}{2024}).
\newblock


\bibitem[Chowdhury et~al\mbox{.}(2021)]%
        {chowdhury2021examining}
\bibfield{author}{\bibinfo{person}{Farhan~Asif Chowdhury},
  \bibinfo{person}{Dheeman Saha}, \bibinfo{person}{Md~Rashidul Hasan},
  \bibinfo{person}{Koustuv Saha}, {and} \bibinfo{person}{Abdullah Mueen}.}
  \bibinfo{year}{2021}\natexlab{}.
\newblock \showarticletitle{Examining factors associated with twitter account
  suspension following the 2020 us presidential election}. In
  \bibinfo{booktitle}{\emph{Proceedings of the 2021 IEEE/ACM international
  conference on advances in social networks analysis and mining}}.
  \bibinfo{pages}{607--612}.
\newblock


\bibitem[Clayton(2022)]%
        {james2022twitter}
\bibfield{author}{\bibinfo{person}{James Clayton}.}
  \bibinfo{year}{2022}\natexlab{}.
\newblock \showarticletitle{Twitter moves to limit Russian government
  accounts}.
\newblock \bibinfo{journal}{\emph{BBC}} (\bibinfo{date}{5 April}
  \bibinfo{year}{2022}).
\newblock
\newblock
\shownote{Available at: \url{https://www.bbc.co.uk/news/technology-60992373}
  (Accessed: {June 5th, 2025})}.


\bibitem[Diskin(2024)]%
        {diskin2024use}
\bibfield{author}{\bibinfo{person}{Evgeni~I Diskin}.}
  \bibinfo{year}{2024}\natexlab{}.
\newblock \showarticletitle{The use of Artificial Intelligence technologies by
  internet platforms for the purposes of censorship}.
\newblock \bibinfo{journal}{\emph{RUDN Journal of Law}} \bibinfo{volume}{28},
  \bibinfo{number}{3} (\bibinfo{year}{2024}), \bibinfo{pages}{584--603}.
\newblock


\bibitem[Elmas(2023)]%
        {elmas2023analyzing}
\bibfield{author}{\bibinfo{person}{Tu{\u{g}}rulcan Elmas}.}
  \bibinfo{year}{2023}\natexlab{}.
\newblock \showarticletitle{Analyzing activity and suspension patterns of
  twitter bots attacking turkish twitter trends by a longitudinal dataset}. In
  \bibinfo{booktitle}{\emph{Companion Proceedings of the ACM Web Conference
  2023}}. \bibinfo{pages}{1404--1412}.
\newblock


\bibitem[Elmas et~al\mbox{.}(2021)]%
        {elmas2021dataset}
\bibfield{author}{\bibinfo{person}{Tu{\u{g}}rulcan Elmas},
  \bibinfo{person}{Rebekah Overdorf}, {and} \bibinfo{person}{Karl Aberer}.}
  \bibinfo{year}{2021}\natexlab{}.
\newblock \showarticletitle{A dataset of state-censored tweets}. In
  \bibinfo{booktitle}{\emph{Proceedings of the International AAAI Conference on
  Web and Social Media}}, Vol.~\bibinfo{volume}{15}.
  \bibinfo{pages}{1009--1015}.
\newblock


\bibitem[Geissler et~al\mbox{.}(2023)]%
        {geissler2023russian}
\bibfield{author}{\bibinfo{person}{Dominique Geissler},
  \bibinfo{person}{Dominik B{\"a}r}, \bibinfo{person}{Nicolas Pr{\"o}llochs},
  {and} \bibinfo{person}{Stefan Feuerriegel}.} \bibinfo{year}{2023}\natexlab{}.
\newblock \showarticletitle{Russian propaganda on social media during the 2022
  invasion of Ukraine}.
\newblock \bibinfo{journal}{\emph{EPJ Data Science}} \bibinfo{volume}{12},
  \bibinfo{number}{1} (\bibinfo{year}{2023}), \bibinfo{pages}{35}.
\newblock


\bibitem[Gosztonyi et~al\mbox{.}(2025)]%
        {gosztonyi2025public}
\bibfield{author}{\bibinfo{person}{Gergely Gosztonyi},
  \bibinfo{person}{J{\'a}nos B{\'a}lint}, {and} \bibinfo{person}{Gergely~Ferenc
  Lendvai}.} \bibinfo{year}{2025}\natexlab{}.
\newblock \showarticletitle{Public Figures and Social Media from a Freedom of
  Expression Viewpoint in the Recent US and EU Jurisdiction}.
\newblock \bibinfo{journal}{\emph{Journalism and Media}} \bibinfo{volume}{6},
  \bibinfo{number}{1} (\bibinfo{year}{2025}), \bibinfo{pages}{26}.
\newblock


\bibitem[Guerin(2025)]%
        {orla2025turkey}
\bibfield{author}{\bibinfo{person}{Orla Guerin}.}
  \bibinfo{year}{2025}\natexlab{}.
\newblock \showarticletitle{Turkey moves to silence jailed Erdogan rival by
  blocking account on X}.
\newblock \bibinfo{journal}{\emph{BBC}} (\bibinfo{date}{8 May}
  \bibinfo{year}{2025}).
\newblock
\newblock
\shownote{Available at: \url{https://www.bbc.co.uk/news/articles/cvgve4q99d5o}
  (Accessed: {June 5th, 2025})}.


\bibitem[Hickey et~al\mbox{.}(2023)]%
        {hickey2023auditing}
\bibfield{author}{\bibinfo{person}{Daniel Hickey}, \bibinfo{person}{Matheus
  Schmitz}, \bibinfo{person}{Daniel Fessler}, \bibinfo{person}{Paul~E
  Smaldino}, \bibinfo{person}{Goran Muric}, {and} \bibinfo{person}{Keith
  Burghardt}.} \bibinfo{year}{2023}\natexlab{}.
\newblock \showarticletitle{Auditing Elon Musk’s impact on hate speech and
  bots}. In \bibinfo{booktitle}{\emph{Proceedings of the international AAAI
  conference on web and social media}}, Vol.~\bibinfo{volume}{17}.
  \bibinfo{pages}{1133--1137}.
\newblock


\bibitem[Jaidka et~al\mbox{.}(2023)]%
        {jaidka2023silenced}
\bibfield{author}{\bibinfo{person}{Kokil Jaidka}, \bibinfo{person}{Subhayan
  Mukerjee}, {and} \bibinfo{person}{Yphtach Lelkes}.}
  \bibinfo{year}{2023}\natexlab{}.
\newblock \showarticletitle{Silenced on social media: the gatekeeping functions
  of shadowbans in the American Twitterverse}.
\newblock \bibinfo{journal}{\emph{Journal of Communication}}
  \bibinfo{volume}{73}, \bibinfo{number}{2} (\bibinfo{year}{2023}),
  \bibinfo{pages}{163--178}.
\newblock


\bibitem[Jain et~al\mbox{.}(2025)]%
        {jain2025follower}
\bibfield{author}{\bibinfo{person}{Pranjal Jain}, \bibinfo{person}{Pooja Jain},
  {and} \bibinfo{person}{Anju Jain}.} \bibinfo{year}{2025}\natexlab{}.
\newblock \showarticletitle{The Follower Fallacy: Revisiting Engagement
  Hypothesis by Evidencing Nonlinear Dynamics in Influencer Marketing.}
\newblock \bibinfo{journal}{\emph{DLSU Business \& Economics Review}}
  \bibinfo{volume}{34}, \bibinfo{number}{2} (\bibinfo{year}{2025}).
\newblock


\bibitem[Lamensch(2024)]%
        {lamensch2024digital}
\bibfield{author}{\bibinfo{person}{Marie Lamensch}.}
  \bibinfo{year}{2024}\natexlab{}.
\newblock \bibinfo{booktitle}{\emph{Digital authoritarianism: The role of
  legislation and regulation}}.
\newblock \bibinfo{publisher}{Centre for International Governance Innovation}.
\newblock


\bibitem[Pierri et~al\mbox{.}(2023)]%
        {pierri2023does}
\bibfield{author}{\bibinfo{person}{Francesco Pierri}, \bibinfo{person}{Luca
  Luceri}, \bibinfo{person}{Emily Chen}, {and} \bibinfo{person}{Emilio
  Ferrara}.} \bibinfo{year}{2023}\natexlab{}.
\newblock \showarticletitle{How does Twitter account moderation work? Dynamics
  of account creation and suspension on Twitter during major geopolitical
  events}.
\newblock \bibinfo{journal}{\emph{EPJ Data Science}} \bibinfo{volume}{12},
  \bibinfo{number}{1} (\bibinfo{year}{2023}), \bibinfo{pages}{43}.
\newblock


\bibitem[Robinson(2024)]%
        {robinson2024moderated}
\bibfield{author}{\bibinfo{person}{Jessica~Yarin Robinson}.}
  \bibinfo{year}{2024}\natexlab{}.
\newblock \showarticletitle{The moderated war in Ukraine: Twitter, Elon Musk,
  and the role of private platforms in war coverage}.
\newblock In \bibinfo{booktitle}{\emph{Media, Dissidence and the War in
  Ukraine}}. \bibinfo{publisher}{Routledge}, \bibinfo{pages}{76--97}.
\newblock


\bibitem[Rohlinger et~al\mbox{.}(2023)]%
        {rohlinger2023does}
\bibfield{author}{\bibinfo{person}{Deana~A Rohlinger}, \bibinfo{person}{Kyle
  Rose}, \bibinfo{person}{Sarah Warren}, {and} \bibinfo{person}{Stuart
  Shulman}.} \bibinfo{year}{2023}\natexlab{}.
\newblock \showarticletitle{Does the Musk Twitter takeover matter? Political
  influencers, their arguments, and the quality of information they share}.
\newblock \bibinfo{journal}{\emph{Socius}}  \bibinfo{volume}{9}
  (\bibinfo{year}{2023}), \bibinfo{pages}{23780231231152193}.
\newblock


\bibitem[Tanash et~al\mbox{.}(2015)]%
        {tanash2015known}
\bibfield{author}{\bibinfo{person}{Rima~S Tanash}, \bibinfo{person}{Zhouhan
  Chen}, \bibinfo{person}{Tanmay Thakur}, \bibinfo{person}{Dan~S Wallach},
  {and} \bibinfo{person}{Devika Subramanian}.} \bibinfo{year}{2015}\natexlab{}.
\newblock \showarticletitle{Known unknowns: An analysis of twitter censorship
  in turkey}. In \bibinfo{booktitle}{\emph{Proceedings of the 14th ACM Workshop
  on Privacy in the Electronic Society}}. \bibinfo{pages}{11--20}.
\newblock


\bibitem[Titcomb(2023)]%
        {james2023putins}
\bibfield{author}{\bibinfo{person}{James Titcomb}.}
  \bibinfo{year}{2023}\natexlab{}.
\newblock \showarticletitle{Putin’s Twitter account resurfaces as Russia
  comes in from the cold}.
\newblock \bibinfo{journal}{\emph{The Telegraph}} (\bibinfo{date}{7 April}
  \bibinfo{year}{2023}).
\newblock
\newblock
\shownote{Available at:
  \url{https://www.telegraph.co.uk/technology/2023/04/07/elon-musk-twitter-lifts-restrictions-putin-kremlin-russia/}
  (Accessed: {June 5th, 2025})}.


\bibitem[Toraman et~al\mbox{.}(2022)]%
        {toraman2022blacklivesmatter}
\bibfield{author}{\bibinfo{person}{Cagri Toraman}, \bibinfo{person}{Furkan
  {\c{S}}ahinu{\c{c}}}, {and} \bibinfo{person}{Eyup~Halit Yilmaz}.}
  \bibinfo{year}{2022}\natexlab{}.
\newblock \showarticletitle{Blacklivesmatter 2020: an analysis of deleted and
  suspended users in Twitter}. In \bibinfo{booktitle}{\emph{Proceedings of the
  14th ACM Web Science Conference 2022}}. \bibinfo{pages}{290--295}.
\newblock


\bibitem[Varol(2016)]%
        {varol2016spatiotemporal}
\bibfield{author}{\bibinfo{person}{Onur Varol}.}
  \bibinfo{year}{2016}\natexlab{}.
\newblock \showarticletitle{Spatiotemporal analysis of censored content on
  twitter}. In \bibinfo{booktitle}{\emph{Proceedings of the 8th ACM Conference
  on Web Science}}. \bibinfo{pages}{372--373}.
\newblock


\bibitem[Voinea(2022)]%
        {voinea2022taking}
\bibfield{author}{\bibinfo{person}{Dan~Valeriu Voinea}.}
  \bibinfo{year}{2022}\natexlab{}.
\newblock \showarticletitle{Taking over Twitter--Balancing Free Speech and
  Content Moderation}.
\newblock \bibinfo{journal}{\emph{Annals of the University of Craiova for
  Journalism, Communication and Management}} \bibinfo{volume}{8},
  \bibinfo{number}{1} (\bibinfo{year}{2022}), \bibinfo{pages}{139--144}.
\newblock


\bibitem[Volkova and Bell(2017)]%
        {volkova2017identifying}
\bibfield{author}{\bibinfo{person}{Svitlana Volkova} {and}
  \bibinfo{person}{Eric Bell}.} \bibinfo{year}{2017}\natexlab{}.
\newblock \showarticletitle{Identifying effective signals to predict deleted
  and suspended accounts on twitter across languages}. In
  \bibinfo{booktitle}{\emph{Proceedings of the International AAAI Conference on
  Web and Social Media}}, Vol.~\bibinfo{volume}{11}. \bibinfo{pages}{290--298}.
\newblock


\bibitem[Wilcoxon(1992)]%
        {wilcoxon1992individual}
\bibfield{author}{\bibinfo{person}{Frank Wilcoxon}.}
  \bibinfo{year}{1992}\natexlab{}.
\newblock \showarticletitle{Individual comparisons by ranking methods}.
\newblock In \bibinfo{booktitle}{\emph{Breakthroughs in statistics: Methodology
  and distribution}}. \bibinfo{publisher}{Springer}, \bibinfo{pages}{196--202}.
\newblock


\end{thebibliography}

\end{document}